\documentclass[11pt,a4paper]{article}
\usepackage{cite,url}
\usepackage{amsmath, amsthm, amssymb,slashed}
\usepackage{ifpdf}
\ifpdf
  \usepackage[pdftex]{graphicx}
  \usepackage{epstopdf}
\else
  \usepackage[dvips]{graphicx}
\fi
\parskip=\baselineskip
\def\Bbb{\mathbb}
\def\Tr{{\rm Tr}}
\def\16{{\bf 16}}
\def\1{{\bf 1}}
\def\2{{\bf 2}}
\def\3{{\bf 3}}
\def\4{{\bf 4}}
\def\be{\begin{equation}}
\def\G{{\mathcal G}}
\def\ee{\end{equation}}

\def\bar{\overline}

\def\R{{\Bbb{R}}}\def\Z{{\Bbb{Z}}}

\def\J{{\mathcal J}}
\font\teneurm=eurm10 \font\seveneurm=eurm7 \font\fiveeurm=eurm5
\newfam\eurmfam
\textfont\eurmfam=\teneurm \scriptfont\eurmfam=\seveneurm
\scriptscriptfont\eurmfam=\fiveeurm

\font\teneusm=eusm10 \font\seveneusm=eusm7 \font\fiveeusm=eusm5
\newfam\eusmfam
\textfont\eusmfam=\teneusm \scriptfont\eusmfam=\seveneusm
\scriptscriptfont\eusmfam=\fiveeusm

\font\tencmmib=cmmib10 \skewchar\tencmmib='177
\font\sevencmmib=cmmib7 \skewchar\sevencmmib='177
\font\fivecmmib=cmmib5 \skewchar\fivecmmib='177
\newfam\cmmibfam
\textfont\cmmibfam=\tencmmib \scriptfont\cmmibfam=\sevencmmib
\scriptscriptfont\cmmibfam=\fivecmmib

\numberwithin{equation}{section}

\def\d{\mathrm d}
\def\C{{\Bbb C}}
\def\Z{{\Bbb Z}}

\def\bar{\overline}

\begin{document}
\begin{titlepage}
\begin{flushright}

\end{flushright}
\vskip 1.5in
\begin{center}
{\bf\Large{An SYK-Like Model Without  Disorder}}
\vskip
0.5cm {Edward Witten} \vskip 0.05in {\small{ \textit{School of
Natural Sciences, Institute for Advanced Study}\vskip -.4cm
{\textit{Einstein Drive, Princeton, NJ 08540 USA}}}
}
\end{center}
\vskip 0.5in
\baselineskip 16pt
\begin{abstract}
Making use of known facts about ``tensor models,'' it is possible to construct a quantum system without quenched disorder that has
 the same large $n$ limit for its correlation functions and thermodynamics as the SYK model. This might be useful in further probes of this
approach to holographic duality.
\end{abstract}
\date{October, 2016}
\end{titlepage}
\def\Hom{\mathrm{Hom}}
\def\F{{\mathcal F}}
\def\i{{\mathrm i}}
\def\Tr{{\mathrm{Tr}}}
\def\CS{{\mathrm{CS}}}
\def\O{{\mathrm O}}
\def\d{{\mathrm d}}
\def\i{{\mathrm i}}
\def\LG{{\mathcal L}G}
\def\R{{\Bbb R}}
\def\C{{\Bbb C}}
\def\Z{{\Bbb Z}}
\def\bar{\overline}
\def\be{\begin{equation}}
\def\ee{\end{equation}}

\section{Introduction}

The SYK model \cite{SY,K} is a quantum mechanical model of fermions with random couplings.
The most widely studied version of the model has 
 $N$ real fermions $\psi_i$, $i=1,\dots,N$, with
$q$-fold random couplings, for some even integer $q\geq 4$.  The model can be described by the action
\be\label{SYK} I=\int \d t \left(\frac{\i}{2}\sum_i\psi_i\frac{\d}{\d t} \psi_i -\i^{q/2}j_{i_1i_2 \dots i_q}
\psi_{i_1}\psi_{i_2}\dots\psi_{i_q}\right). \ee
Here the couplings are Gaussian random variables.  If we write $I$ for a multi-index $i_1i_2\dots i_q$,
the the $j_I$ are drawn from a Gaussian ensemble with variance 
\be\label{variance}\langle j_I j_{I'}\rangle =\delta_{II'}\frac{J^2 (q-1)!}{N^{q-1}}, \ee
for some constant $J$.

\begin{figure}
 \begin{center}
   \includegraphics[width=4.5in]{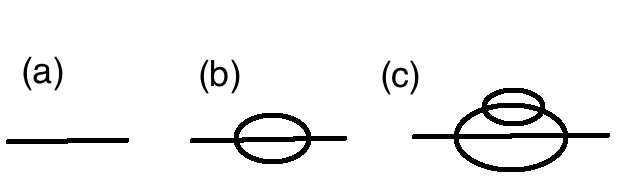}
 \end{center}
\caption{\small The iterative procedure that generates the leading diagrams of the SYK model
for large $N$. Figures are drawn for $q=4$ (quartic vertices).   At each step of the process, one replaces
a propagator (part (a)) with the two-loop 
diagram of part (b).  For example, after another iteration, one can generate the 
four-loop diagram of part (c).}
\label{Iteration}
\end{figure}

The model is solvable in the limit of large $N$, fixed $J$ by a saddle point method that reflects
the fact that the dominant Feynman diagrams in this limit have a simple type.   They can be generated
by an iterative procedure that is illustrated in fig. \ref{Iteration}.  At each stage of the iteration, one takes
a propagator (fig \ref{Iteration}(a)) and replaces it by the diagram of fig. \ref{Iteration}(b).  This
can be done any number of times to generate more complicated diagrams.  For example, at the second
step of the iteration, one can generate the diagram of fig. \ref{Iteration}(c).

A model of this general type was formulated many years ago \cite{SY} to describe a spin-fluid state, but the subject
has attracted renewed interest because of the suggestion \cite{K} that the large $N$ limit of the model
is dual to a black hole in an emergent $1+1$-dimensional spacetime.  This type of model was first discussed
in relation to holographic duality in \cite{S}.   For recent work, see 
\cite{S2,HQRY,PR,YLX,FS,ASY,MS,BAK,JS,GQS,GrR,GV,FGMS,BA,BNRS}.

One question that one might ask about the SYK model is whether, instead of formulating it as a model
with quenched disorder, one can find similar physics in a more conventional large $N$ limit.
One idea might be to interpret the couplings $j_I$ as quantum variables with very slow dynamics, rather
than random constants.  Even if it works, this has the drawback that the thermodynamic entropy
of the $j_I$ (roughly $N^q/q!$ bosonic variables) will overwhelm that of the $\psi_i$ ($N$ fermionic
variables).  

The present paper is devoted to another approach that does not have this drawback.  In fact, there is an already
known class of ``tensor models'' whose large $N$ limits are governed by precisely the same Feynman diagrams
that dominate the large $N$ limit of the SYK model.     There have been many papers on this subject, a sampling
being \cite{G1,GR,G2,BGRR,BGR,R,G3,G4}.  In this literature, the dominant graphs are called ``melons'' or
``melonic graphs.''  A convenient reference is \cite{BGRR}.  
The construction of tensor
models was motivated by the idea of generalizing to higher dimensions the  familiar relation of matrix
models to random two-dimensional geometries.\footnote{This idea also motivated an earlier literature on more generic
tensor models that do not necessarily have a large $N$ limit similar to that of the SYK model.
See for example \cite{Amb,Gr, Sa}.}
   In that application, the dimension $D$ is related to the
order $q$ of the SYK interactions by $D=q-1$.  The status of this program is unclear,
since it is not clear that the rather special Feynman diagrams that are generated by the procedure summarized
in fig. \ref{Iteration} describe a useful class of random $D$-geometries.  But because the tensor models
are governed in the large $N$ limit by the same Feynman diagrams as the SYK model, they do give a framework 
to eliminate the quenched disorder of the SYK model in favor of a more standard large $N$ limit.

It is not entirely clear that this is helpful, but there are at least two reasons that it might be.  First, the average
of a quantum system over quenched disorder is not really a quantum system, so the reliance on 
quenched disorder might make it difficult to apply the SYK model to some subtle questions about black holes.
Second, in the SYK model, certain bilinear expressions in the $\psi_i$, which roughly speaking are
``singlets'' in a disorder-averaged sense, appear to have natural duals in the emergent two-dimensional world.
But it is not clear that the $N$ fields $\psi_i$ themselves have a natural interpretation of that sort.  At any rate,
their analogs do not have bulk duals in better-understood examples of gauge/gravity duality.
The tensor models that mimic the SYK model can be chosen, as we will do in section \ref{themodel},
to have a global symmetry group
$G$ whose dimension is relatively large, but still much less than $N$.   Gauging the $G$ symmetry
leaves as gauge-invariant operators the singlet operators that are relevant in the gravitational dual description,
but eliminates the elementary fermion fields themselves.  But because the dimension of $G$ is much less
than $N$, gauging the symmetry does not significantly affect the thermodynamics of the model when $N$ is large.

In section \ref{themodel}, we describe a variant of the SYK model that has a similar large $N$ limit but without quenched disorder.
In doing this, we adapt and modify the construction described in \cite{BGRR} (and other papers cited above) 
in minor ways.  Our fields are fermion fields in $0+1$ spacetime dimensions
rather than boson fields in 0 dimensions.  Also, we construct a simple model with a relatively large amount of symmetry
(which could be gauged) and do not discuss some of the more generic models that have been considered in the literature.
None of this affects the counting of powers of $N$ in Feynman diagrams.  As in \cite{K} and many
subsequent papers, we use
real fermion fields.  As a result, the index loops we encounter are unoriented and certain two-manifolds that arise may be unorientable.
We could instead use complex fermion fields as in \cite{SY}; then as in \cite{BGRR}, the index loops and two-manifolds will be oriented.

In section \ref{somedetails}, following \cite{BGRR}, we sketch the proof that the Feynman diagrams 
that survive in the large $N$ limit
of the tensor model are precisely those of the SYK model.   The $1/N$ corrections are different, though the significance of the difference
is not clear.

\section{The Model}\label{themodel}

The model will be constructed with $q=D+1$ real fermion fields $\psi_0,\dots,\psi_D$.  Each will have $n^D$ real components, for some
integer $n$, so the total number of real fermion fields will be $N=(D+1)n^D$.  

For each $a$, the field $\psi_a$ will transform in a real irreducible representation of a symmetry group $G$, as follows.
First of all, for each unordered pair $a,b$ of distinct elements of the finite set $\{0,1,\dots,D\}$, we introduce a copy $G_{ab}$ of the
group $\O(n)$.  Since the pair $a,b$ is unordered, we do not distinguish $G_{ab}$ from $G_{ba}$.  The full symmetry group of the model
is then a product \be\label{torf}G_0=\prod_{a<b}G_{ab}\cong \O(n)^{D(D+1)/2}, \ee  up to a discrete quotient that we consider in a moment.

For each $a$, we now declare that $\psi_a$ transforms as the tensor product of the
 vector representations of $G_{ab}$ for every $b\not=a$, and transforms
trivially under $G_{bc}$ if $b,c\not=a$.  The vector representation of $G_{ab}$ is $n$-dimensional, and there are $D$ groups $G_{ab}$ with
$b\not=a$, so $\psi_a$ has $n^D$ real components, as stated above.    With this choice for the fermion representation, a certain discrete
subgroup of $G_0$ acts trivially.  The center of $\O(n)$ is $\Z_2$, acting by sign change on the vector representation, so the
center of $G_0$ is $\Z_2^{D(D+1)/2}$.  A certain subgroup $\Z_2^{(D-2)(D+1)/2}$ acts trivially on all the $\psi_a$, so the group that
will be a faithfully acting symmetry group of the theory is 
\be\label{symgr}G=G_0/\Z_2^{(D-2)(D+1)/2}.\ee

We now replace the SYK action with 
\be\label{nSYK} I=\int \d t \left(\frac{\i}{2}\sum_i\psi_i\frac{\d}{\d t} \psi_i -\i^{q/2}j
\psi_0\psi_1\dots\psi_D\right), \ee
with a real coupling parameter $j$.  The meaning of the expression $\psi_0\psi_1\dots\psi_D$ is as follows.  For each $a<b$, precisely
two of these fields, namely $\psi_a$ and $\psi_b$, transform as vectors under $G_{ab}$.  Contracting the vector indices of $G_{ab}$ for each pair $a,b$,
we arrive at a $G$-invariant that we denote for brevity as $\psi_0\psi_1\dots\psi_D$.

Note that the Hamiltonian of the theory is $H=\i^{q/2}j
\psi_0\psi_1\dots\psi_D$.  This Hamiltonian is odd under a {\it unitary} transformation that changes the sign of one of the $\psi_a$.
(For a discussion of Hamiltonians with the unusual property of being odd under a unitary symmetry, see \cite{Moore}.)
It is observed in \cite{MS} that the Hamiltonian of the SYK model has this property on a statistical basis. Permutations
of the $\psi_a$ accompanied by corresponding permutations of the $G_{ab}$ are global symmetres (an odd permutation must be accompanied by a sign change of one of the $\psi_a$).

Now we describe some basic properties of Feynman diagrams of this theory.    
In fig. \ref{Comparison}, we depict in two different ways the basic Feynman vertex of the theory for 
$D=3$ or $q=4$.  In part (a), it is  simply shown as a basic quartic vertex for four fields $\psi_0$, $\psi_1$,  $\psi_2$, and $\psi_3$.  Propagation
of one of these fields is represented as usual by a ``line'' in the Feynman graph.  In part (b), we take note of the fact that each of 
the $\psi_a$ is a trifundamental of a certain product of three copies of $\O(n)$, and so can be considered to carry three $n$-valued
``indices.''   Accordingly, we resolve each ``line'' in part (a) into three ``strands'' in part (b).   For example, a line of type 0 is resolved
into strands of type 01, 02, and 03, reflecting the fact that $\psi_0$ transforms as a trifundamental of $G_{01}\times G_{02}\times G_{03}$.
Then to construct the Feynman vertex, we connect all of the strands in the only way consistent with their labeling, arriving at the figure.

   \begin{figure}
 \begin{center}
   \includegraphics[width=5in]{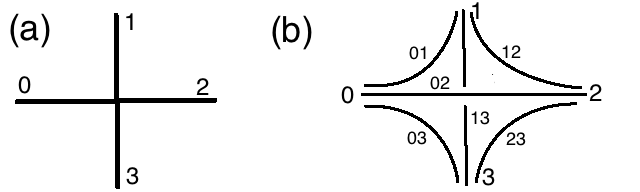}
 \end{center}
\caption{\small Two views of the basic Feynman vertex of the theory.  Here and later, figures are drawn for quartic vertices ($q=4$ or
$D=3$). In (a), the vertex is drawn as a simple quartic
vertex with external lines labeled $0,1,2$, or 3.  In (b), each line is resolved into three ``strands,'' representing how the ``indices''
of a field transform under a symmetry group.  For example, the field $\psi_0$ is a trifundamental of $G_{01}\times G_{02}\times G_{03}$,
so it is represented with three strands labeled $01,$ $02$, and $03$.  The vertex is constructed by connecting these
strands in the only way consistent with their labeling. It can be visualized as a tetrahedron.}
\label{Vertex}
\end{figure}

   \begin{figure}
 \begin{center}
   \includegraphics[width=5in]{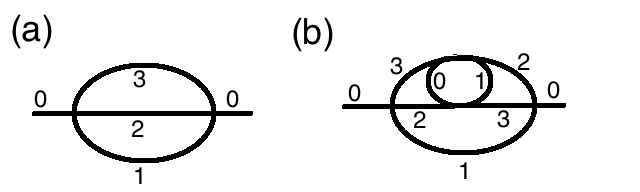}
 \end{center}
\caption{\small A typical diagram that survives (a) or does not survive (b) in the large $n$ limit.}
\label{Comparison}
\end{figure} 

Hopefully the reader can visualize fig. \ref{Vertex}(b) as representing a tetrahedron, with each vertex labeled by an index $a\in\{0,1,2,3\}$
and the edge connecting vertices $a$ and $b$ labeled as $ab$.   This interpretation of the vertex as a tetrahedron is part of the 
relationship of the tensor models described in \cite{G1,GR,G2,BGRR,BGR,R,G3,G4} (for $D=3$) with three-geometries. For larger $D$, the tetrahedron is
replaced by a $D$-simplex and the graphs can be interpreted as $D$-geometries.  However, as noted in the introduction,
the dominant Feynman diagrams, for large $n$, correspond to a fairly special class of $D$-geometries.  The relation to the SYK model
gives an alternative motivation to consider this class of model.

Now we make some preliminary remarks on the large $n$ limit, or equivalently the large $N$ limit.  To count powers of $n$ in a Feynman diagram, it is very convenient to think about the fact that each
line in the diagram represents $D$ strands, as in fig. \ref{Vertex}(b).   The strands can form closed loops and, as in the more familiar
matrix models, such a closed loop gives a factor of $n$.  A difference from matrix models is that there are different kinds of closed
loops.  A strand may be of type $ab$ for any unordered pair $a,b\in \{0,1.\dots,D\}$ and hence there are altogether $D(D+1)/2$ distinct
kinds of strand.  We let $\F_{ab}$ be the number of closed loops made of strands of type $ab$, and 
\be\label{zelf}\F=\sum_{a<b}\F_{ab}. \ee
In \cite{BGRR}, $\F_{ab}$ is called the number of faces of type $ab$ and $\F$ the total number of faces.  (The rationale for this
terminology is that if one glues in a disc whose boundary is a closed strand of type $ab$, one gets a two-dimensional ``face''
of some triangulated geometry.)  Summing over index loops will give a factor of 
\be\label{welf}n^\F=\prod_{a<b}n^{\F_{ab}}.\ee

Let us work this out in some simple examples.  Some simple Feynman diagrams for $D=3$ are drawn in fig. \ref{Comparison}.  The
diagrams are drawn in the standard way, representing the propagation of one of the $\psi_a$ by a line and not resolving the lines
into strands.  The reason is that diagrams soon become rather complicated if drawn in terms of strands.  With a little practice, one
can easily count powers of $n$ without drawing the strands.  Every closed loop in the graph that only contains lines of type $a$
or $b$ will, when resolved in strands, give a closed strand of type $ab$.  This is the only source of closed strands of this type,
so $\F_{ab}$ is just the number of closed loops that can be drawn in the graph using only lines labeled $a$ or $b$.  We call
these loops of type $ab$.  (Loops of type $ab$
are always disjoint, because of the form of the interaction vertex.)   For example, in fig. \ref{Comparison}(a) one can form a single
closed loop of type $12$, $13$, or $23$, and none of type $0a$ for any $a$.  So for this diagram, $\F=\sum_{ab}\F_{ab}=3$.
Thus the diagram is proportional to $j^2n^3$.  So to ensure a large $n$ limit, we must
take
\be\label{pelf} j=\frac{J}{n^{3/2}} \ee
for some constant $J$.  

Once we scale the coupling with $n$ so that fig. \ref{Comparison}(a) has a large $n$ limit, it is almost immediate that any diagram made
by the iterative procedure of fig. \ref{Iteration} likewise has a large $n$ limit.  However, other diagrams vanish for large $n$.
Postponing a systematic explanation for section \ref{somedetails}, we first examine in fig. \ref{Comparison}(b) a simple diagram
that is not generated by the iterative procedure.  This diagram has one closed loop of type 01, one of type 23,
one of type 12, and one of type 13.  With four vertices,
this diagram is of order $j^4n^4\sim 1/n^2\sim 1/N^{2/3}$.  Thus the diagram is subleading in the large
$n$ or equivalently the large $N$ limit.  However, the $1/n$ corrections are different in this model
from what they are in the SYK model, as the expansion in that model is an expansion in integer
powers of $1/N$.

\begin{figure}
 \begin{center}
   \includegraphics[width=4in]{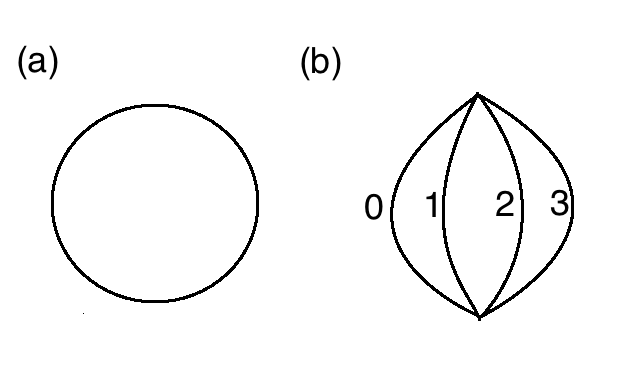}
 \end{center}
\caption{\small (a) The vacuum amplitude in free fermion theory is represented by this one-loop diagram
(the loop may carry any label 0,1,2, or 3). (b)  The lowest
order nontrivial contribution to the vacuum amplitude, obtained by applying the iterative step in fig. \ref{Iteration}
to the one-loop diagram in (a).}
\label{Basic}
\end{figure}

\begin{figure}
 \begin{center}
   \includegraphics[width=3.5in]{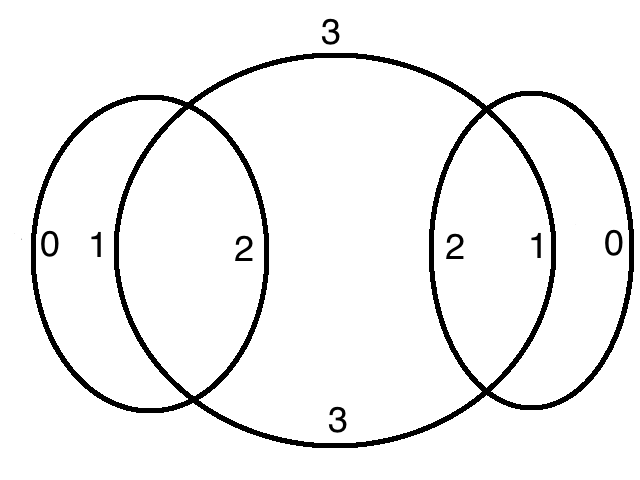}
 \end{center}
\caption{\small The next step in the iteration  can generate this more complicated vacuum diagram, which also survives in the large $N$
limit.}
\label{Vacuum}
\end{figure}

The examples we have considered were contributions to the two-point function.  In a similar way, we can consider
vacuum diagrams.  In free field theory, we would just have the one-loop diagram of fig. \ref{Basic}(a).  With $N$ fermion
fields, it makes a contribution of order $N$.  Applying once the usual iterative step of fig. \ref{Iteration}, we get the first nontrivial contribution
to the vacuum amplitude in fig. \ref{Basic}(b).  It is of order $j^2n^6=J^2N$ and thus is again of order $N$ for large $N$.
Another iterative step can lead, for example, to the diagram of fig. \ref{Vacuum}, which is of order $j^4n^9=J^4N$.
In general, the leading contributions to the vacuum amplitude are generated by the iterative procedure and are of order $N$.

\section{Some Details}\label{somedetails}

Following \cite{BGRR}, we will explain how to analyze the large $n$ behavior of the perturbative expansion in a model of this kind.  (We describe only the leading behavior.  It is not clear how difficult it is to systematically describe the diagrams
that arise in each order in $1/n$.)

It does not matter very much whether we study the large $n$ limit for vacuum diagrams or for correlation functions,
since leading order contributions to the $2k$-point function are made by ``cutting'' $k$ lines in a diagram that makes
a leading order contribution to the vacuum amplitude.  
  Thus, to understand the large $n$ behavior of the perturbation expansion, it essentially suffices
to consider the vacuum amplitude.   This will give a minor simplification in the following.

Let $\J=(a_0,a_1,\dots,a_D)$ be an unoriented cyclic ordering of the set $\{0,1,\dots,D\}$; the term ``unoriented'' means that
  two cyclic orderings that differ by a mirror
reflection are considered equivalent.  Given such a $\J$, we can reduce any Feynman graph  $\G$, such as the graph of fig. \ref{Vacuum},
to the graph of a more familiar matrix model, as follows.   Each line in $\G$ of type $a_i$ (for some $i$) can be resolved
in strands of type $a_ib$ for all $b\not=a_i$, making $D$ strands in all.  In the abstract, there is no natural way to pick just 2 of
the strands associated to a given line.   However, once we are given the cyclic arrangement $\J=(a_0,a_1,\dots,
a_D)$, each label $a_i$ has the two ``neighbors'' $a_{i+1}$ and $a_{i-1}$.  For each line of type $a_i$, we remember only the
two strands of type $a_{i\,i\pm 1}$ and forget the others.  When in this way we keep only two strands for each line in the graph,
the lines we keep always meet smoothly at vertices and $\G$ becomes a ribbon graph or fatgraph of a matrix model.  (This will be a matrix model with unoriented index loops, as our fields are
real and the individual strands are unoriented.)  In the fashion familiar from matrix models, by gluing in discs whose boundaries
are the closed index loops of type $a_{i\,i+1}$ (for all $i$), we make a closed two-manifold that we will call $\Sigma_\J$, since
it depends on $\J$.

An important fact will be that the Euler characteristic of $\Sigma_\J$, which we denote $\chi_\J$, can be no larger than $2$.  We denote
it as
\be\label{zeff} \chi_\J=2-2g_\J,\ee
but we note that as our fields are real, $\Sigma_\J$ may be unorientable, so $\chi_\J$ may be odd and $g_\J$ defined this
way may be a half-integer.  At any rate, the important
property is that $g_\J$ is nonnegative.   Following \cite{BGRR}, we define the ``degree'' of the graph $\G$ as
\be\label{zegree}\omega(\G)=\sum_\J g_\J=\sum_\J\left(1-\frac{\chi_\J}{2}\right).\ee
Thus $\omega(\G)\geq 0$ for all $\G$, and if $\omega(\G)=0$ then $g_\J=0$ for all $\J$, which means that the $\Sigma_\J$ are all
two-spheres.

As an example of the definition  of $g_\J$, we consider the diagram of fig. \ref{Vacuum} and take $\J=(0,1,2,3)$.  This means that we are supposed
to keep strands of type $i\,i+1$ for any $i$.  The planar diagram made this way is simply the obvious planar diagram associated
with the fact that the graph of fig. \ref{Vacuum} has been drawn in a plane.  $\Sigma_\J$ is a two-sphere, constructed by adding
a point at infinity to the plane in which the diagram has been drawn.

Let $v_0$ and $v_1$ be the number of vertices and edges in the graph $\G$.  These do not depend on the choice of $\J$.
However, the number of faces (discs that are glued in when we construct $\Sigma_\J$) does depend on $\J$.  We denote
this number as $v_{2,\J}$.  It is
\be\label{orf} v_{2,\J}=\sum_{i=0}^D \F_{a_i a_{i+1}},\ee
since  the  faces of $\Sigma_\J$ are associated to index loops of type $a_i a_{i+1}$ (for some $i$).
We also have $v_1=\frac{D+1}{2}v_0$, because $\G$ is constructed from $(D+1)$-valent vertices.

The Euler characteristic of $\Sigma_\J$ is 
\be\label{chiz}\chi_\J=v_0-v_1+v_{2,\J} =-\left(\frac{D-1}{2}\right)v_0+\sum_i \F_{a_i a_{i+1}}. \ee
From this formula and (\ref{zegree}), we can work out a useful formula for $\omega(\G)$:
\be\label{pliz}\frac{2}{(D-1)!}\omega(\G)=D+\frac{D(D-1)}{4}v_0 -\F. \ee
The main point in the derivation is that each pair $ab$ are adjacent in precisely $(D-1)!$ of the orderings $\J$ and hence each
$\F_{ab}$ occurs $(D-1)!$ times when eqn. (\ref{chiz}) is summed over $\J$.  

Now we define the large $n$ limit by taking 
\be\label{priz}j=\frac{J}{n^{D(D-1)/4}}\ee
with fixed $J$.  A Feynman graph $\G$ with $v_0$ vertices and $\F$ closed strands (of any type $ab$) will be proportional then
to 
\be\label{orfit} n^{-(D(D-1)/4)v_0+\F}=n^{D-2\omega(\G)/(D-1)!}=N^{1-2\omega(\G)/D!}. \ee
Since $\omega_\G\geq 0$ for all $\G$, the large $N$ limit of any graph is at most proportional to $N$,
and the graphs that do make contributions of order $N$ are precisely those -- such as that of fig. \ref{Vacuum} -- with
$\omega_\G=0$.

It remains then to understand which graphs do have $\omega_\G=0$.  The basic statement here (Lemma 1 in \cite{BGRR}) is
that any graph with $\omega_\G=0$ has a face of some type $ab$ with only two vertices (in other words, some strand of type $ab$
forms a closed loop after
passing through only two vertices; for example, in fig. \ref{Vacuum} there are two strands of type 01, two of type 02, and two of type 12 with
this property). The proof in \cite{BGRR} is as follows.

If $\omega_\G=0$, by definition this means that the total number of faces is $\F=\frac{D(D-1)}{4}v_0+D$.  Given the structure of the interaction
vertex $\psi_0\psi_1\dots\psi_D$, in which each field appears only once, each closed strand must pass through an even number of vertices.
Write $\F_s$ for the number of faces with $2s$ vertices, so
\be\label{plick}\F_1+\F_2+\sum_{s>2}\F_s=\F=\frac{D(D-1)}{4}v_0+D. \ee
Let $2p_\rho^{ab}$ be the number of vertices of the $\rho^{th}$ closed strand (or face) of type $ab$.  One-half the total number of vertices of any face is
\be\label{rlick}\sum_{\rho,a<b}p_\rho^{a,b}=\F_1+2\F_2+\sum_{s>2}s\F_s.\ee
Each vertex contributes to $D(D+1)/2$ faces of some type, so $\sum_{\rho,a<b}p_\rho^{a,b}=D(D+1)v_0/4$.   Combining these formulass and eliminating
$\F_2$, we get
\be\label{qlick}\F_1=2D+\sum_{s\geq 3}(s-2)\F_s +\frac{D(D-3)}{4}v_0. \ee
Thus $\F_1>0$ for $D\geq 3$.  For the SYK model, one has $q\geq 4$ and thus automatically $D=q-1$ is $\geq 3$.  (However, there
are variants of the SYK model derived from random cubic tensors \cite{FGMS}, and it is not immediately apparent how to express large $N$ limits of these
models without quenched randomness.)

    \begin{figure}
 \begin{center}
   \includegraphics[width=4.5in]{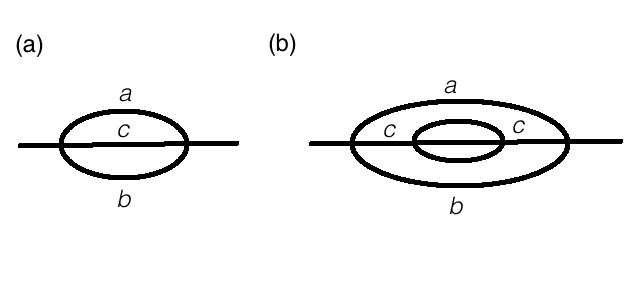}
 \end{center}
\caption{\small If $\G$ has degree 0 and a face $\F_*$  of type $ab$ with only two faces, then a planar diagram of any type $(\dots acb\dots )$
looks near $\F_*$ like (a) or (b).  In (a), the two vertices of $\F_*$ are connected by a single line labeled
$c$  and
in (b) this line is replaced by a more complicated graph such as the one shown.}
\label{Choices}
\end{figure} 
Suppose now that the graph $\G$ has a face $\F_*$ of type $ab$ with precisely 2 vertices.   As noted earlier, if the degree $\omega(\G)$ vanishes, then for any cyclic order $\J$, the two-manifold 
$\Sigma_\J$ is topologically a sphere, and the graph corresponding to $\J$ is planar.
 Let $c$ be any label in the set $\{0,1,\dots,D\}$ other
than $a$ and $b$.  
  Consider a cyclic order $\J$ that reads in part $(\dots acb\dots)$ (in other words, part of the sequence is $acb$).     Near $\F_*$, the
planar graph that corresponds to the ordering $\J$ will have to
look like part (a) or (b) of fig. \ref{Choices} (this is essentially fig. 2 of \cite{BGRR}).   The two cases differ by whether the two vertices of $\F_*$ are
connected by a single propagator with label $c$ (part (a) of the figure) or something more complicated (part (b)).

If the picture looks like fig. \ref{Choices}(a) for all $c\not=a,b$, then the graph $\G$ is precisely the diagram of fig. \ref{Basic}(b),
which arises at the first nontrivial step of the iteration that produces the leadng graphs of the SYK model.  

    \begin{figure}
 \begin{center}
   \includegraphics[width=2.5in]{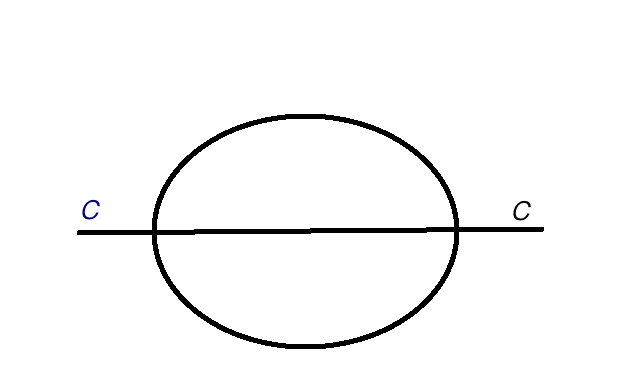}
 \end{center}
\caption{\small Given a graph $\G$ as in fig. \ref{Choices}(b) that makes a leading contribution to the vacuum amplitude, the ``interior'' of the
face $\F_*$ is a graph that makes a leading contribution to the two-point function. This graph is shown here for the example
of fig. \ref{Choices}(b).  Gluing together the external lines of ths
graph, we get another graph $\G'$, with fewer vertices than $\G$, that makes a leading order contribution to the vacuum amplitude.
In the case shown, this graph is simply the basic one of fig. \ref{Basic}(b).}
\label{Interior}
\end{figure} 

Suppose instead that the picture looks like fig. \ref{Choices}(b) for some $c$.  Then we finish the argument by an induction on the number $n_0$ of
vertices in the graph $\G$.    Suppose a graph $\G$ that
contributes a leading order term to the vacuum amplitude has a face $\F_*$ with two vertices with a local picture that looks
like fig. \ref{Choices}(b).  Then the ``interior'' of the face $\F_*$ is a graph (fig. \ref{Interior}) that makes a leading order contribution to the two-point function.
If we glue together its two external lines, we get a graph $\G'$ that makes a leading order contribution to the vacuum
amplitude.  (In the case shown in fig. \ref{Interior}, this will just be again the basic diagram of fig. \ref{Basic}(b).)
On the other hand, we can make another graph $\G''$ that also makes a leading order contribution to the vacuum
amplitude by replacing the interior of $\F^*$ by a single propagator (in other words, we modify $\G$ by locally replacing
fig. \ref{Choices}(b) by fig. \ref{Choices}(a)).  Both $\G'$ and $\G''$ have fewer vertices than $\G$ so by the inductive
hypothesis, we can assume that they are each generated starting with the one-loop vacuum diagram of fig. \ref{Basic}(a)
by the inductive procedure of the SYK model.  But then the same
is true for $\G$.

To explain part of this more intuitively, we make the following remark.  If a graph $\G$ has the property that $g_\J=0$ for some
$\J$, then it is a planar diagram and can be drawn on a two-sphere.  But if $\omega(\G)=0$, then $g_\J=0$ for all $\J$
and there are many different ways to draw $\G$ on a two-sphere.  A generic planar diagram can be drawn on a two-sphere in essentially
only one way.  The inductive procedure that generates the leading order diagrams of the SYK model generates
diagrams that can be drawn on the two-sphere in as many ways as possible.  For example, fig. \ref{Basic}(b) is drawn as a planar
diagram, and after any permutation of the labels 0,1,2, and 3 it is still planar.

\vskip.2cm
Research supported in part by NSF Grant PHY-1606531.  I thank J. Maldacena, D. Stanford, and J. Suh for discussions.

\bibliographystyle{unsrt}

\end{document}